\documentclass[apjl]{emulateapj}
\usepackage{apjfonts}

\newcommand{\um}{${\rm \mu m}$~}
\newcommand{\mm}{${\rm \mu m}$}

\long\def\symbolfootnote[#1]#2{\begingroup%
\def\thefootnote{\fnsymbol{footnote}}\footnote[#1]{#2}\endgroup}

\slugcomment{accepted for publication in ApJ.} 

\begin{document}

\title{The Debris Disk around HR 8799}
 \author{K. Y. L. Su\altaffilmark{1}, 
   G. H. Rieke\altaffilmark{1}, 
   K. R. Stapelfeldt\altaffilmark{2},
   R. Malhotra\altaffilmark{3},
   G. Bryden\altaffilmark{2},
   P. S. Smith\altaffilmark{1},
   K. A. Misselt\altaffilmark{1},
   A. Moro-Martin\altaffilmark{4,5},
   J. P. Williams\altaffilmark{6}
  }

\altaffiltext{1}{Steward Observatory, University of Arizona, 933 N
  Cherry Ave., Tucson, AZ 85721; ksu@as.arizona.edu}
\altaffiltext{2}{JPL/Caltech, 4800 Oak Grove Drive, Pasadena, CA
  91109}
\altaffiltext{3}{Lunar and Planetary Laboratory, University of Arizona}
\altaffiltext{4}{Department of Astrophysical Science, Princeton
  University, USA}
\altaffiltext{5}{Center for Astrobiology (CSIC-INTA), Madrid, Spain}
\altaffiltext{6}{Institute for Astronomy, University of Hawaii, USA} 
\begin{abstract}

We have obtained a full suite of {\it Spitzer} observations to
characterize the debris disk around HR 8799 and to explore how its
properties are related to the recently discovered set of three massive
planets orbiting the star. We distinguish three components to the
debris system: (1) warm dust (T $\sim$ 150 K) orbiting within the
innermost planet; (2) a broad zone of cold dust (T $\sim$ 45 K) with a
sharp inner edge, orbiting just outside the outermost planet and
presumably sculpted by it; and (3) a dramatic halo of small grains
originating in the cold dust component. The high level of dynamical
activity implied by this halo may arise due to enhanced
gravitational stirring by the massive planets. The relatively young
age of HR 8799 places it in an important early stage 
of development and may provide some help in understanding the
interaction of planets and planetary debris, an important process in
the evolution of our own solar system.

\end{abstract} 

\keywords{circumstellar matter -- infrared: stars -- planetary systems
-- stars: individual (HR8799)}

\section{Introduction}

Rapid progress has been made in the past decade in discovering other
planetary systems and in building theories for their formation and
evolution. More than 300 extrasolar planets have been found through
measuring stellar radial velocities and about 60 planets have been
observed to transit their stars (Extrasolar Planets Encyclopaedia,
http://exoplanet.eu/). However, these techniques are optimal in
finding systems dramatically different from the solar system, with
giant planets in orbits very close to their stars.  There are two
primary avenues for probing systems with planets that might have
structures more analogous to that of the solar system. First, giant
planets have recently been imaged around Fomalhaut \citep{kalas08} and
HR 8799 \citep{marois08} in orbits with radii of tens of AU. Second,
the Spitzer Space Telescope ({\it Spitzer}, \citealt{werner04}) has
observed more than 200 debris disks, which represent dust production
from colliding planetesimals, typically in zones analogous to the
asteroid and Kuiper belts in the solar system. It is expected that
these debris disks are sculpted by giant planets through gravitational
interactions (Wyatt 2008 and references therein). Observations and
models of well-studied examples show a wide variety of structures and
behavior, possibly related to the individual quirks of their (unseen)
planetary systems and their evolutionary states (e.g.,
\citealt{stapelfeldt04,su05,beichman05,song05,rhee08,su08,hillenbrand08,backman09}).

HR 8799 has been known to have a prominent debris system presumably
sculpted by unseen planets \citep{zuckerman04}.  The recent detection
of three giant planets orbiting this star, along with estimates of
their masses and orbits, provides a new opportunity to understand in
more detail the forces that shape debris disks. This paper reports new
measurements with {\it Spitzer} that fully characterize the debris
disk around HR 8799. We show that the debris system includes at least
three components: warm dust interior to the three planets; cold dust
exterior to them; and a halo of small grains surrounding the cold dust
zone. The placement of the first two disk components appears to be as
expected from the orbits of the planets.  The dynamics of the three
massive planets \citep{fabrycky08,reidemeister09,goz09} suggests
substantial dynamical excitation in the overall planetary system.  The
elevated level of dynamical activity in the cold dust zone, which is
the source of the large halo, is possibly a result of the unstable
state of the system. The 20--160 Myr age of the system
\citep{moor06,marois08} corresponds to a critical phase in the history
of our own solar system when planet formation processes were near
completion and the solar system's dynamical configuration was being
established.  The HR 8799 debris system offers an exciting view of
this phase of planetary system formation and evolution.

We present the new observations in \S 2, describe the general
observational results in \S 3, build detailed models in \S 4, and
discuss the implications of this work in \S 5.

\section{Observations and Data Reduction} 
\label{obs}

The observations presented here are from the {\it Spitzer} Director's
Discretionary Time (DDT) program 530 and Guaranteed Time Observations
(GTO) program 50175 that utilize all of the major observation modes of
both the InfraRed Spectrograph (IRS,
\citealt{houck04}) and the Multiband Imaging Photometer for {\it Spitzer}
(MIPS, \citealt{rieke04}). The
observational setting and depth 
for each of the modes are summarized in Table \ref{obs_tab}.  A
short (6 s $\times$ 1 cycle) integration IRS spectrum of HR 8799 was
previously published by \citet{chen06,chen09}. Here we present an 
IRS spectrum that is 4--60 times deeper along with newly obtained MIPS images. 
Observations at 24 $\mu$m were obtained in standard
small-field photometry mode, for a total integration of 210 s. The
observation at 70 $\mu$m provides a total integration of $\sim$320 s
on source. The 160 \um observation was obtained using the
enhanced-mode at 4 sub-pixel-offset positions, for a total integration
of $\sim$150 s.  The MIPS SED-mode observation provides a
low-resolution (R=15--25) spectrum from 55 to 95 \um with 600 s of
integration on source.
 
\begin{figure*}
 \figurenum{1}
 \label{fig1}
 \plotone{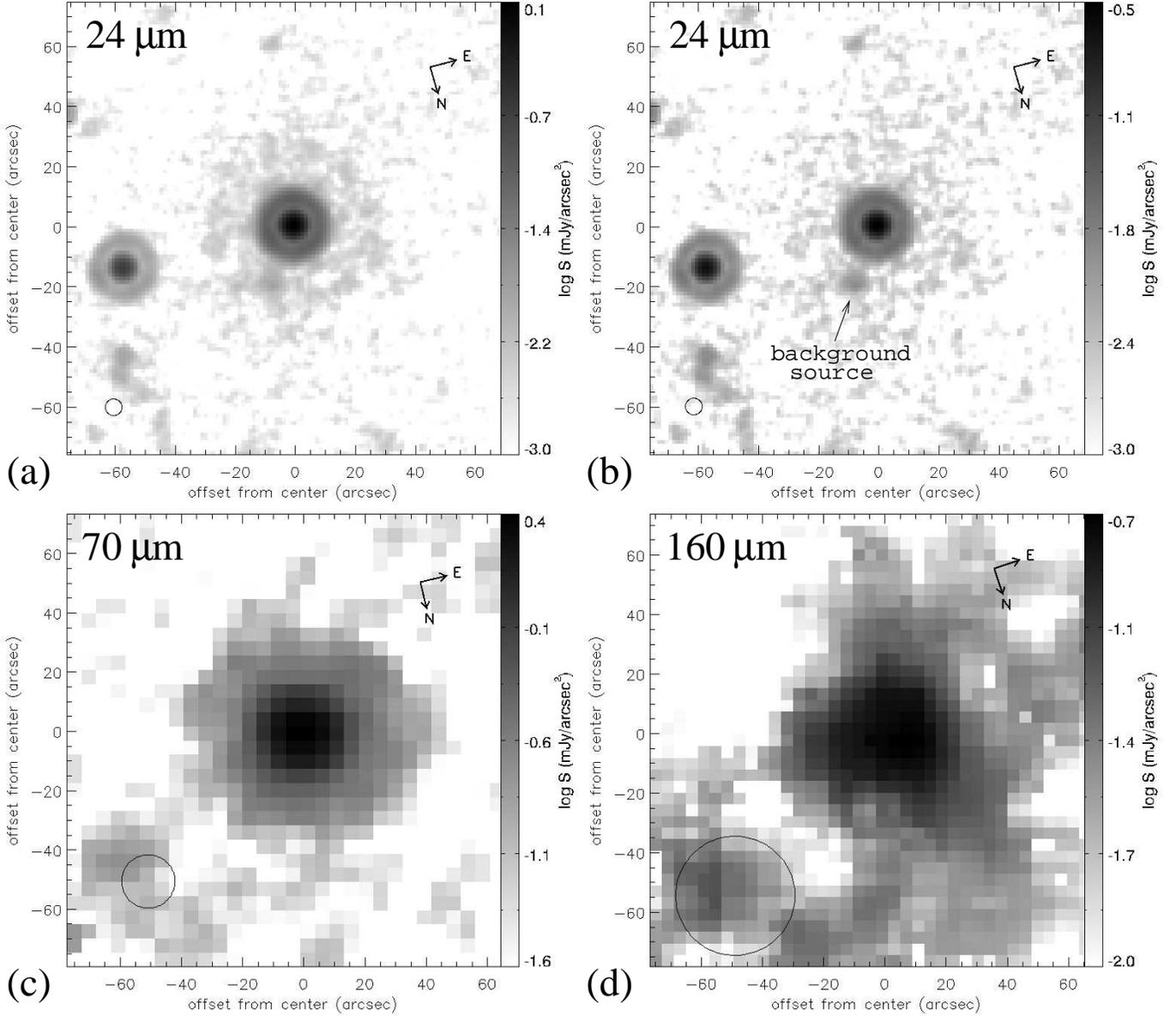} 
\caption{MIPS images of HR 8799 with the orientation and beam sizes
  (FWHMs, black circles) indicated and the surface brightness scale on the
  side of each panel. (a) 24 \um image before photospheric
  subtraction. (b) 24 \um image after subtraction of the model stellar
  photosphere. (c) and (d) are the 70 and 160 \um images that are
  dominated by emission from the disk. The faint 24 \um background source
  located $\sim$21\arcsec~from HR 8799 is marked by a black arrow. The
  bright 24 \um source located $\sim$60\arcsec~away is a field G-type
  star, BD+20 5278p. } 
 \end{figure*}

\begin{deluxetable}{rlll}
\tablewidth{0pc}
\tablecaption{{\it Spitzer} Observations of HR 8799\label{obs_tab}}
\tablehead{ 
\colhead{AOR Key} & \colhead{Instrument} & \colhead{Module} &
\colhead{Integration} 
}
\startdata
28889856   &  IRS Staring & SL2     &   6 s $\times$ 4 cycles \\
           &              & SL1     &  14 s $\times$ 4 cycles \\
           &              & LL2     & 120 s $\times$ 3 cycles \\
           &              & LL1     & 120 s $\times$ 3 cycles \\
28889088   &  MIPS Photometry & 24 \mm, 5 cluster pos. & 3 s $\times$ 1 cycle\\
28889344   &  MIPS Photometry & 70 \mm, default scale & 10 s $\times$ 4 cycles\\
28889600   &  MIPS Photometry &160 \mm, 4 cluster pos.&  10s $\times$ 1 cycle\\
           &                  &  (enhanced mode)      &           \\ 
25711872   &  MIPS SED-mode   &  1\arcmin~chop  &  10s $\times$10 cycles \\
\enddata 
\end{deluxetable}

All of the MIPS data were processed using the Data Analysis Tool
\citep{gordon05} for basic reduction (e.g., dark subtraction, flat 
fielding/illumination corrections), with additional processing to
minimize instrumental artifacts \citep{engelbracht07,gordon07}.  After
correcting these artifacts in individual exposures, the final mosaics
were combined with pixels half the size of the physical pixel scale;
the resulting images are shown in Figure \ref{fig1}. The calibration
factors used to transfer the instrumental units to the physical unit
(mJy) are adopted from the MIPS calibration papers
\citep{engelbracht07, gordon07, stansberry07, lu08}.

To ensure the best flat field result at 24 \mm, a second flat field
made from a median stack of the data with all bright sources masked
out was also applied in addition to the regular scan-mirror-dependent
flat fielding (for details see \citealt{engelbracht07}). The flux
density of the source at 24 \um was estimated using aperture
photometry with both small and large apertures (the same photometry
parameters\symbolfootnote[1]{small aperture: a radius of 6\farcs23 with sky
annulus from 19\farcs92 to 29\farcs88 and an aperture correction of
1.699; large aperture: a radius of 14\farcs94 with sky annulus from
29\farcs88 to 42\farcs33 and an aperture correction of 1.142.} 
as used
in \citealt{su06}). The large aperture gives a flux density of
83.0$\pm$0.5 mJy, while the small aperture gives a flux density of
80.8$\pm$0.7 mJy ($\sim$2.7\% lower), implying the source profile is
more extended than a true point source. Therefore, to capture all the
flux from the extended component, we conducted photometry with a very
large (radius of 35\arcsec) aperture on an image with the DC offset
and all background sources subtracted without a sky
annulus (i.e., the aperture photometry was relative to the median
pixel value for the entire frame; therefore, an aperture correction of
1.06 was applied, see \citealt{engelbracht07}). The final measured
flux we adopted is 86.4 mJy (before color correction).

The center of the source at 24 \um was
determined using a 2-D Gaussian fitting routine, and it coincides
with the expected stellar position within the pointing error
($<$1\arcsec). This is consistent with the fact that the flux
contribution in the 24 \um band is mostly from the stellar photosphere
(58 mJy; see \S \ref{photosphere}).  
The stellar photosphere was then 
subtracted by scaling an observed blue point spread function
(PSF). After photospheric subtraction (see Fig.~\ref{fig1}b), the
source has FWHMs of 5\farcs78$\times$5\farcs67 at a position angle
(P.A.) of 17\arcdeg\symbolfootnote[2]{The disk brightness in the 24 \um band
  is dominated by the bright unresolved component (details see \S
  \ref{analysis:disk_components}). Therefore, the
  position angle given here only reflects the angle of the instrumental
  PSF, not the extended disk.}, slightly broader than
an observed red PSF (5\farcs61$\times$5\farcs55 for the $\zeta$ Lep
disk, \citealt{su08}). There is a faint ($\sim$0.8 mJy) source
located $\sim$21\arcsec~from HR 8799 at P.A. of --36\arcdeg, presumably
a background object.

The 70 \um data reduction follows the steps recommended by
\citet{gordon07} and uses time filtering with the source region masked
out. Several region sizes were tried. A masked radius of
55\arcsec~yields the minimum value for the background variation
(1-$\sigma_{70}$ = 1.14$\times$10$^{-2}$ mJy~arcsec$^{-2}$ per
subpixel). The source at 70 \um (see Fig.~\ref{fig1}c) is clearly
extended with FWHMs of 24\farcs6$\times$24\farcs4 at a P.A. of
77\arcdeg\symbolfootnote[3]{Note that source appears to be
azimuthally symmetric, therefore the P.A. quoted here does not
necessarily reflects the true P.A. of the source.}, compared to the
nominal resolution of 18\arcsec. The source has an azimuthally
symmetric morphology in the image.  The outer boundary of the source
(though masked by the instrumental PSF) can be traced out to
42\arcsec~or 34\arcsec~in radius at 1- or 3-$\sigma_{70}$ levels,
respectively.  Aperture photometry was used to estimate the flux
density for the source since it is extended. An aperture of
42\arcsec~in radius (1-$\sigma$ boundary) was used with a sky annulus
of 44\arcsec--54\arcsec. After correcting for lost light based on the
same aperture settings applied to a theoretical PSF, the final
integrated 70 \um flux density is 545 mJy (before color correction).

The 160 \um data were taken in the ``enhanced AOT'' to allow them to
be time filtered, as was also done for the 70 \um data (for details
see \citealt{stansberry07}). No additional 160 \um PSF was obtained
for the purpose of leak subtraction as it has been shown that the
ghost image produced by the 160 \um filter leakage is less than
$\sim$15 times of the photospheric flux density at 160 \mm. The disk is
$\sim$400 times brighter than the photosphere in this channel (see
Tab.~\ref{tbl:fluxes}). This is also confirmed by careful inspection
of the data where the expected ghost image lies. The final 160 \um
mosaic shows some large-scale extended cirrus structures surrounding
the source (see discussion in \S \ref{bkg_cirrus}). The source is
clearly detected at 160 \um (see Fig.~\ref{fig1}d), and is consistent
with being azimuthally symmetric above the 5-$\sigma_{160}$ level 
at the expected star position. 
The source at 160 \um is surrounded by a low level of asymmetric
background cirrus. 
The FWHMs are 47\arcsec$\times$41\arcsec~(based
on an image with a field of view of 88\arcsec), slightly broader
than a nominal observed PSF ($\sim$38\arcsec, \citealt{stansberry07}),
possibly influenced by the background cirrus. 
Aperture photometry is used to determine the integrated
flux. To minimize the influence of the background cirrus, we used
small aperture sizes (16\arcsec~and 24\arcsec), sky annuli
(64\arcsec--128\arcsec) and aperture corrections (4.683 and 2.613 for
a red point source of 50 K; \citealt{stansberry07}). The
resultant integrated 160 \um flux density is 539 mJy (before
color-correction). The centroid of the disk 
seen in the MIPS 70 and 160 \um images coincides with the expected
position of the star. 

\begin{deluxetable}{rrrrrrr}
\tablewidth{0pc}
\tablecaption{Observed Flux Densities \label{tbl:fluxes}}
\tablehead{ 
\colhead{$\lambda_c$} & \multicolumn{6}{c}{Flux Density (mJy)} \\
\colhead{$\mu$m}      & \colhead{Total\tablenotemark{a}}   &
\colhead{Star\tablenotemark{b}}
&\colhead{Disk} & \colhead{ MIR Core\tablenotemark{c}} & \colhead{
  FIR Core\tablenotemark{d}} 
& \colhead{Extended\tablenotemark{e}}  
}
\startdata
 23.68 & 86.6  & 58 & 29 &  21       &   $<$5  & 7.5 \\ 
 71.42 & 610 &  6.5 & 605  &  \nodata  & 285--387& 319--217 \\ 
155.89 & 555 &  1.4 & 554  &  \nodata  & 554     & \nodata \\
\enddata
\tablenotetext{a}{Integrated in a large aperture including the star
  after color correction}
\tablenotetext{b}{From best-fit Kurucz model}
\tablenotetext{c}{Central unresolved source for the inner warm
  component at 24 \um}
\tablenotetext{d}{Central unresolved source for the outer cold
  component at 70 \um} 
\tablenotetext{e}{Extended emission for the cold component}

\end{deluxetable}

The MIPS SED-mode data were reduced and calibrated as described by
\citet{lu08} with an extraction aperture of 5 native pixels
($\sim$50\arcsec) in the spatial direction. Since the source size (see
below) is smaller than the extraction aperture, the slit loss was
corrected based on a point source. This procedure would under-correct
for the slightly extended source and may explain why there is small
offset ($\sim$17\% at the fiducial wavelength of the 70 \um band)
between the 70 $\mu$m photometry and the MIPS-SED spectrum. The final
MIPS SED-mode spectrum is smoothed to match the resolution at the
long-wavelength portion of the spectrum (R=15).

\begin{figure}
\figurenum{2}
\label{sed}
\plotone{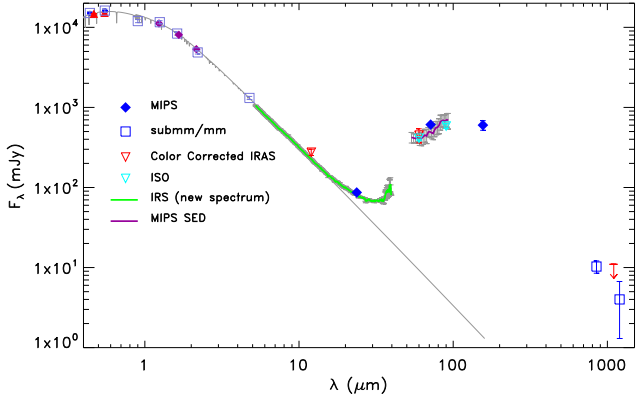}
\caption{Spectral Energy Distribution (SED) of HR 8799. The best-fit
  Kurucz model (T$_{eff}$=7750 K with log $g$ of 4.5, shown as a grey thin line) was
  determined using ground-based photometry for wavelengths shorter
  than 6 \um compiled from the Simbad database (including the 2MASS
  photometry shown as purple filled diamonds, Str\"omgen photometry shown as red filled triangles,
  Hipparcos $BV$ photometry as blue filled circles, and various
  Johnson photometry shown as open squares). The ISO 60 and 90 \um measurements are from
  \citet{moor06} while the 850 \um datum is from \citet{williams06} and the
  1.1 and 1.2 mm observations are from \citet{sylvester96}. The MIPS
  broad-band measurements are color corrected. }
\end{figure}

The basic reduction and extraction of the IRS spectral data were
provided by the SSC IRS pipeline S18.5.~We first trimmed a few end
points from each module and manually removed large outliers by
visually inspecting each of the co-added, background-subtracted
spectra (differencing the two nod positions).  The spectra were then
averaged with sigma clipping (S/N=3). All of the final photometry
measurements (color-corrected, see details in \S
\ref{analysis:disk_components}), MIPS SED-mode data, IRS combined
spectrum, along with previously published data are shown in the
spectral energy distribution (SED) (Fig.~\ref{sed}).

\section{Analysis} 
\label{analysis}

\subsection{Stellar Properties of HR 8799}
\label{photosphere}

HR 8799 (HD~218396, HIP114189), located at 39.4 pc
\citep{vanleeuwen07}, is classified as an A5 V \citep{cowley69},
$\gamma$ Doradus variable. It has $\lambda$ Bootis type
characteristics \citep{gray99} with low abundances of the heavier
elements (e.g., [Fe/H] = --0.55), but near-solar abundances of C and O
(Sadakane 2006).  A high-resolution optical spectrum of the star
indicates $T_{eff}$ = 7250K, log $g$ = 4.30 (Sadakane 2006). The
galactic space motion (UVW) of the star resembles those of young
clusters and associations in the solar neighborhood with ages of
20--160 Myr \citep{moor06,marois08}. The stellar rotation velocity ($v
sini$) is 49 km~s$^{-1}$ \citep{royer07}, consistent with the star
being viewed close to pole-on.

To determine the stellar spectral energy distribution, 
we fit all available optical to near-infrared
photometry (Johnson $UBV$, Str\"omgen $uvby$ photometry, Hipparcos Tycho
$BV$ photometry, 2MASS $JHK_s$ photometry) with the synthetic Kurucz
model \citep{castelli03} based on a $\chi^2$ goodness of fit test. 
A value of $T_{eff}$=7500 K with $R_{\ast}$=1.4 $R_{\sun}$ and log g =
4.5 and sub-solar abundances gives the best
match, in good agreement with the results from spectroscopy (Sadakane 2006). 
The resulting stellar luminosity is 5.7 $L_{\sun}$, placing the star near
the zero-age main sequence consistent with a young age. Based on the 
best-fit Kurucz model, we then estimate the stellar photospheric flux densities at
23.6, 71.42, and 155.9 \um for the MIPS 24, 70 and 160 \um bands,
respectively. These values are listed in Table \ref{tbl:fluxes}. 
With a stellar mass of 1.5 $M_{\sun}$, the blowout size ($a_{bl}$) is
$\sim$2 \um assuming a grain density of 2.5 g~cm$^{-3}$ around HR 8799.

\begin{figure}
\figurenum{3}
\label{excesssed}
\plotone{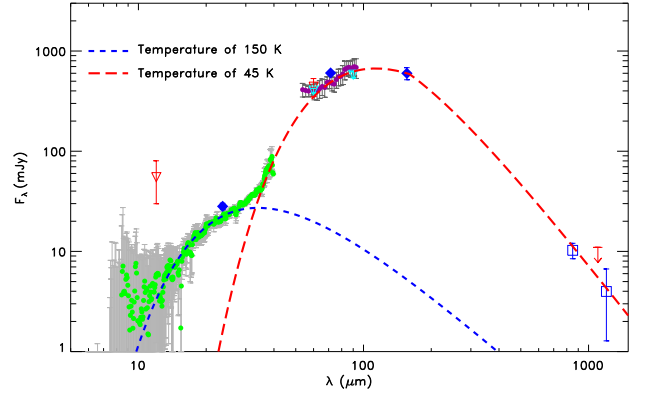}
\caption{SED of the HR 8799 Disk (after stellar photospheric
  subtraction). The symbols used are the same as in Figure \ref{sed}. The
short dashed blue line shows the blackbody emission at 150 K that
matches the excess shortward 30 \um well, while the excess longward 30
\um is fit well with modified  blackbody emission
($\lambda^{-0.95} B_{\lambda}$) at 45 K (long dashed red line). The MIPS 24
and 70 \um photometry points include the extended component; therefore
they are slightly higher than the spectra.}
\end{figure}

\subsection{Disk Components} 
\label{analysis:disk_components} 

\begin{figure}
\figurenum{4}
\label{radius_td}
\plotone{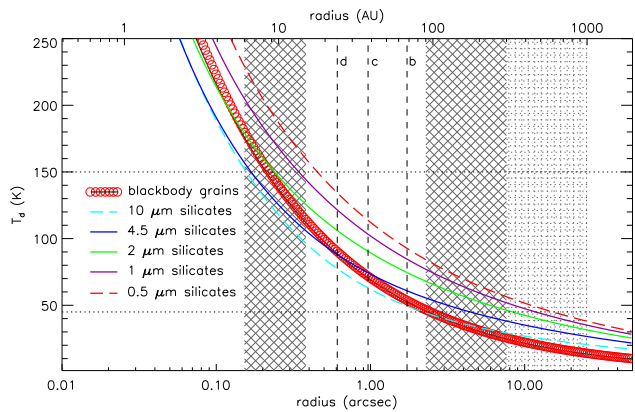}
\caption{Thermal equilibrium dust temperature in the HR 8799
  system computed based on various grain sizes of astronomical silicates. The two
  horizontal dotted lines mark the dust temperatures 
  (150 and 45 K) derived from the excess emission for blackbody radiators. For
  illustration, the projected distances of the three planets 
  are marked as vertical dashed lines. The three disk components are marked in
  different filled patterns (see \S \ref{analysis:disk_components} for
  details).}
\end{figure}

Figure \ref{excesssed} shows the excess SED after stellar photospheric
subtraction. It is evident that there are at least two disk components
in the system: one with a characteristic temperature of $\sim$150 K
(warm) and the other with a characteristic temperature of $\sim$45 K
(cold). Similar disk components have been suggested by
\citet{reidemeister09} as well.  
For blackbody radiators that are in thermal equilibrium with
the stellar radiation field, the warm (150 K) and cold (45 K)
components correspond to radii of $\sim$9 AU (0\farcs2) and $\sim$95
AU (2\farcs4) from the star. Interestingly, HR 8799 d, c
and b are at projected separations of 24, 38 and 68 AU
from the star, respectively \citep{marois08}.  
They all lie between the two dust reservoirs.
In general, the location of the dust
that is at a specific thermal equilibrium temperature depends on the
grain properties (through the absorption efficiency $Q_{abs}$, which
is size dependent). The equilibrium dust temperature is computed by
balancing the absorption and emission energy and ultimately it depends
on cross sections ($Q_{abs}\pi a^2$, where $a$ is the grain
radius); therefore, 
the temperature of large grains is generally lower than that of small
grains at the same distance from the heating source. We adopt the
optical constants for astronomical silicates \citep{laor93}
and compute the grain properties ($Q$ values for the absorption and
scattering efficiency for grain radii of 0.1--1000 \um in size) using
Mie theory with additional modifications for large sizes 
($a\gtrsim$10 \mm). The resultant thermal equilibrium dust
temperatures as a function of radius from the star are shown in
Figure \ref{radius_td}.  
Due to the nature of an optically thin debris system, any dust in a
continuous spatial distribution between the two distinct temperature
components will have intermediate temperatures, resulting in a smoother
excess SED between 20--55 \um (see Fig.~2 in
\citealt{reidemeister09}). The SED shape and the distinct 
characteristic dust temperatures strongly suggest that very little
dust resides between the two components. 

Based on Figure \ref{radius_td}, the emitting zone for the warm
component is between $\sim$5 and $\sim$15 AU, with an angular diameter
of $<$ 1\arcsec.  This component, which dominates the emission
detected in the IRS spectrum and the MIPS 24 \um band, should be
unresolved at 24 \um (inside planet~d). A similar calculation
indicates that the cold component, if it has a size of $\sim$100 AU in
radius, should be at best only barely resolved at 24 \mm; however, the
excess SED (Fig.~\ref{excesssed}) shows that the cold component would
contribute only a small ($\lesssim$ 9\%) portion of the total signal in the
MIPS 24 \um band. Thus, the flux well outside the PSF at 24 \um suggests the 
detection of an extended component at this wavelength.

We estimate the flux of the unresolved point source in the 24 \um
image (star + disk) by scaling a PSF to match the peak flux of the
source. The scaling suggests that the central unresolved component is
$\sim$78 mJy at maximum (assuming the extended component contributes
no flux at the position of the star). The stellar photosphere is
expected to be 58 mJy in the 24 \um band, so the central unresolved
disk component is $\sim$20 mJy before color correction. Due to the
large temperature differences between the warm and cold components,
different color-corrections must be applied to each of the
components at 24 \um. Applying a color-correction factor of 1.055 for
a blackbody of 150 K (see the MIPS Data Handbook,
http://ssc.spitzer.caltech.edu/mips/dh/, for the detailed color
corrections for the MIPS bands), the total flux density of the central
unresolved (warm) disk component is $\sim$21 mJy, which agrees well
with the photosphere-subtracted IRS spectrum (22.4$\pm$1.4
mJy). Therefore, the extended disk component is $\sim$7.5 mJy (after
applying a color correction of 0.894 for a blackbody of 50 K). 
Adding these measurements, the
final total flux density for the entire system (star + two disk
components after color corrections for the two different temperatures)
is 86.6 $\pm$1.7 mJy (assumed 2\% error based on the nominal
calibration uncertainty, \citealt{engelbracht07}).

The expected stellar photosphere is 6.5 mJy in the 70 \um band,
suggesting the disk is $\sim$603 mJy after a color correction factor
of 1.12 for a 50 K blackbody. Therefore, the total (star + disk) flux
density at 70 \um is 610$\pm$31 mJy (assumed 5\% error based on the
nominal calibration uncertainty, \citealt{gordon07}). At 160 \mm,
the stellar photosphere is 1.4 mJy; therefore, the total disk flux
density is $\sim$554 mJy after a color correction of 1.03 for a
blackbody of 50 K, resulting in a total (star+disk) flux of 555$\pm$66
mJy (assumed 12\% error based on the nominal calibration uncertainty,
\citealt{stansberry07}).

\begin{figure}
\figurenum{5}
\label{radprof70}
\plotone{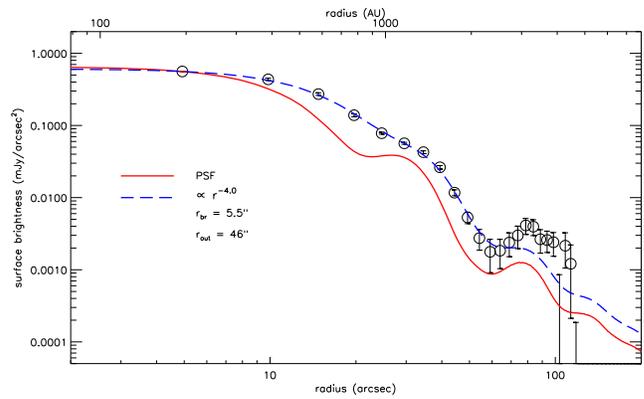}
\caption{Radial surface brightness profile of the HR 8799 disk at 70
  \mm. Data are shown in open black circles with error bars compared
  to the profile of a PSF (red solid line) scaled to match the peak of
the disk flux. An extended (outer radius of 46\arcsec) model surface
brightness profile is also shown as a dashed blue line for
comparison (For details, see \S
\ref{analysis:surface_brightness_distribution}). The profile at radii
of 75\arcsec--90\arcsec~is where the second bright Airy ring of the
instrumental PSF falls, and the match in the profiles between
observed point sources and the theoretical PSF is poor in this range. } 
\end{figure}

The 70 \um image is essentially only from the disk since the star is
$\lesssim$1 \% of the total flux in the far-infrared. An observed
radial surface brightness 
profile of the disk at 70 \um is shown in Figure \ref{radprof70}. As
was indicated by the FWHM measurements, the disk is clearly resolved.
The structure of the disk at 70 \um is similar to the one at 24 \mm, a
bright core and an extended disk; however, the bright core at 70 \um
cannot originate from the warm component seen at 24 \um (because the
temperature is too cold). This bright core is most likely to originate
from a dust belt (hereafter the planetesimal disk) outside planet~b,
where emission by dust is also detected at submillimeter wavelengths
\citep{williams06,sylvester96}. Consistent with this possibility, the
emission at 850 \um is mostly confined within 20\arcsec~(diameter) based on
the JCMT/SCUBA measurements \citep{williams06}. However, if
the dust emitting at 70 \um all originated from the same location as in
the submillimeter, the disk could not be resolved by MIPS at 70 \um. 
In addition, \citet{reidemeister09} also estimate the outer edge of the
cold dust component is between 125 and 170 AU (a narrow ring) based on
simple SED fitting and assumed grain properties. Given the nominal
resolution of 18\arcsec~in the MIPS 70 \um band, any disk structure that is
smaller than 360 AU in radius is not resolvable at a distance of
40 pc. The large measured FWHMs at 70 \um suggest that the extended
emission at 70 \um must originate from a component outside the planetesimal disk.

Limited by the large beam size at 70 \mm, the current data cannot
differentiate whether there is any gap between the planetesimal disk
and the extended component. We can only estimate the bright core
(presumably the planetesimal disk) flux
by normalizing a PSF to match the peak of the observed flux (this
procedure provides an upper limit) or to match the flux within a
10\arcsec~radius (to provide a lower limit). After color correction
for 50 K, the unresolved core flux density is 285--387 mJy, suggesting
that the extended component is roughly equal in integrated brightness
to the unresolved core.
 
\subsection{Disk Inclination}

Based on 10 years of astrometric data on planet b,
\citet{lafreniere09} suggest an orbital inclination of
13\arcdeg--23\arcdeg~off the plane of the sky. In addition, both
\citet{fabrycky08} and \citet{reidemeister09} suggest that the system
is unlikely to be exactly face-on from stability analyses of the
planetary configuration. The resolved image at 24 \um provides very
little constraint because the emission at this wavelength is mostly
dominated by the central unresolved disk. The resolved image at 70 \um
appears to be relatively azimuthally symmetric. From $\sim$100
MIPS observations of a routine calibrator, HD~180711 (G9 III, 447 mJy
at 70 \mm), the measured FWHM 
ratio is 1.044$\pm$0.023 for a point source at 70 \mm. The measured FWHM
ratio (1.008) of the HR 8799 70 \um disk suggests a deviation of
$\sim$1.5-$\sigma$ from a face-on 
symmetric disk. 
From model images (see discussion in \S \ref{analysis:thermal_model})
constructed for various inclination angles, we can rule out any angles
that are larger than $\sim$25\arcdeg~because the resultant images give
a ratio much larger than the measured value.

\subsection{Summary} 

In summary, the resolved disks at 24 and 70 \um indicate that the disk
around HR 8799 has at least three components: a warm belt inside
planet~d, a planetesimal disk outside of planet~b with extension of
$\lesssim$10\arcsec, and a surrounding large extended disk. The separation 
into these three components and their placement relative to the
planets in the system appear to be largely independent of assumed
grain sizes and optical properties. All of the observed flux densities,
including those for the three disk components, are listed in Table
\ref{tbl:fluxes}.

\begin{figure}
\figurenum{6}
\label{radprof24}
\plotone{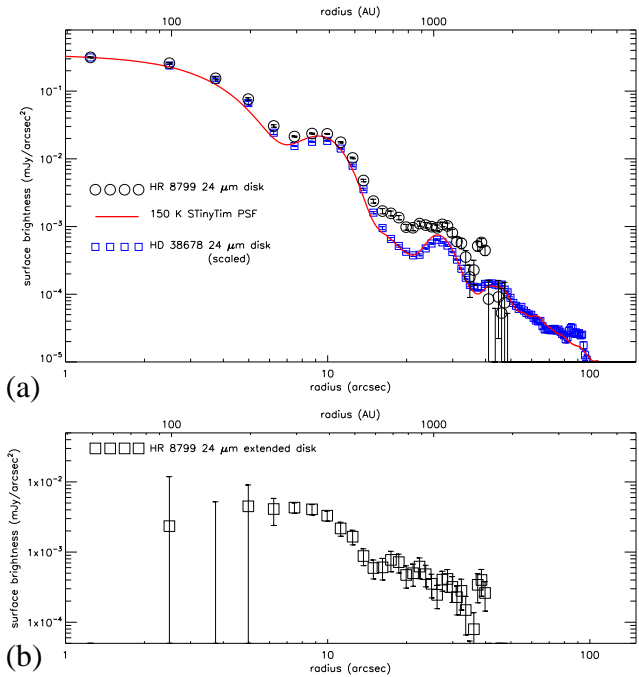} 
\caption{(a) radial profile cut of the HR 8799 disk at 24 \um (black
open circles). For comparison, similar radial profiles (scaled to
match the peak of the HR 8799 disk flux) are shown based on the
theoretical STinyTim PSF (boxcar smoothed, 150 K) and the unresolved
disk around HD~38678 ($\zeta$ Lep). The unresolved disk around
HD~38678 has high signal-to-noise and shows a dynamical range of
2$\times$10$^4$ in the radial profile (Su et al.  in prep.). An
extended disk filling the second and third dark Airy rings is evident
in the profile for HR 8799. (b) radial profile cut of this extended
disk at 24 \um after subtracting the scaled $\zeta$ Lep disk
(color-corrected).}
\end{figure}

\section{Models}
\label{sec_models}

We now build on the observational results of the preceding section
through more detailed models of the HR 8799 debris disk. Although
these specific models are under-constrained and hence not unique,
they are guided by models of other debris systems and fit 
the observations of HR 8799.  Therefore, they give a plausible picture
of the disk behavior. In
\S~\ref{analysis:surface_brightness_distribution}, we constrain the
disk extent by fitting the observed surface brightness profiles with
assumed powerlaw profiles that are independent of grain
properties. The main purpose of this exercise is to derive spatial
information that can be used in a realistic thermal model.  In
\S~\ref{analysis:thermal_model}, we then utilize the disk components
that are inferred from the observations with the adopted grain
properties to estimate the physical extent and masses required for
the dust in each component.

\subsection{Disk Surface Brightness Distributions} 
\label{analysis:surface_brightness_distribution}

Because of our low angular resolution, it is convenient to
characterize the disk using radially averaged surface brightness
distributions, represented as a mean value at a given radius and the
standard deviation of the mean. For simplicity, we assume that the
disk is face-on, so an azimuthally averaged radial profile can be used for
modeling. This approximation greatly reduces the noise by averaging
the intensity at a given radius (a similar approach has been used on
the disks around Vega \citep{su05} and $\epsilon$ Eri
\citep{backman09}). The surface brightness radial profile of the disk
(after PSF subtraction for the stellar photosphere and masking out all
nearby sources) at 24 \um is shown in Figure
\ref{radprof24}a. For comparison, we also show the azimuthally
averaged radial profiles of the smoothed theoretical PSF (generated by the
STinyTim program, \citealt{krist06}) and the unresolved disk around
$\zeta$~Lep\symbolfootnote[4]{Most of the available 24 \um PSF calibrators
  are stars and thus have very blue colors compared to the HR 8799
  disk. We use the profile of the $\zeta$~Lep
disk to demonstrate the accuracy of our data reduction and the
stability of the {\it Spitzer} PSF, by matching the theoretical radial
profile over a range in brightness of over more than 4 orders of magnitude.}
(Su et al.~2009, in prep.) scaled 
to match the peak flux of the HR 8799 disk. The radial profile of the
unresolved $\zeta$ Lep disk agrees well with the theoretical PSF,
while it is clear that the second dark and the third bright Airy 
rings of the HR 8799 disk (in the range of $\sim$20\arcsec~to
$\sim$30\arcsec~from the center) are brighter than for a point source.

By subtracting the scaled radial profile of the $\zeta$ Lep disk from
the observed radial profile of HR 8799, a rough disk surface brightness
distribution for the extended disk component is revealed
(Fig.~\ref{radprof24}b). This component is very faint,
$\sim$1$\times$10$^{-3}$ mJy arcsec$^{-2}$ at a distance of
$\sim$1000 AU from the star, similar to the surface brightness level
seen at 24 \um in the Vega disk.

As shown in Figures \ref{radprof70} and \ref{radprof24}, the true disk
structure is masked by the instrumental PSFs. One way to understand
the structure of the disk is to construct a trial surface brightness
distribution, convolve it with the beam, and then compare with the
observed data. In particular, this approach is a practical way to
constrain the true disk outer radius in the low-resolution images.
Since our goal is to set constraints on the main planetesimal disk,
the 24 \um profile in the following fits is for the extended disk
only.  The simplest model is to assume that the surface brightness
profile follows a power-law, $S(r)\sim r^{-\alpha}$, between an inner
break radius ($r_{br}$) and outer cut-off radius ($r_{out}$). In an
optically thin debris disk where the only heating source is the
stellar radiation and the disk density is only a function of radius in
a power-law form of $\sim r^{-p}$, the radial surface brightness is
then proportional to $r^{-p} B_{\lambda}(T_r)$ where $B_{\lambda}$ is
the Planck function, and $T_r$ is the radial-dependent dust temperature
(generally, $T_r \sim r^{-0.33}$ for small grains and $T_r \sim
r^{-0.5}$ for large grains).  
This simple power-law formula has been used to fit the well-resolved Vega
disk at both 24 and 70 \mm, and it was found that the power
indices change from $r^{-3}$ for the inner part of the disk to
$r^{-4}$ for the outer part of the disk \citep{su05}. 
An inner break point is needed in this simple
power-law for steep indices ($\alpha \sim$3 or 4); otherwise, the
profile mimics a point source after convolution with the PSF. Because
the outermost planet is at a radius of $\sim$2\arcsec, we also set the
disk surface brightness to zero inside 2\arcsec, i.e., $S(r)=$0 for
$r\le r_h=$2\arcsec~($h$ stands for $hole$).

\begin{figure*} 
\figurenum{7}
\label{radprof_fits} 
\plotone{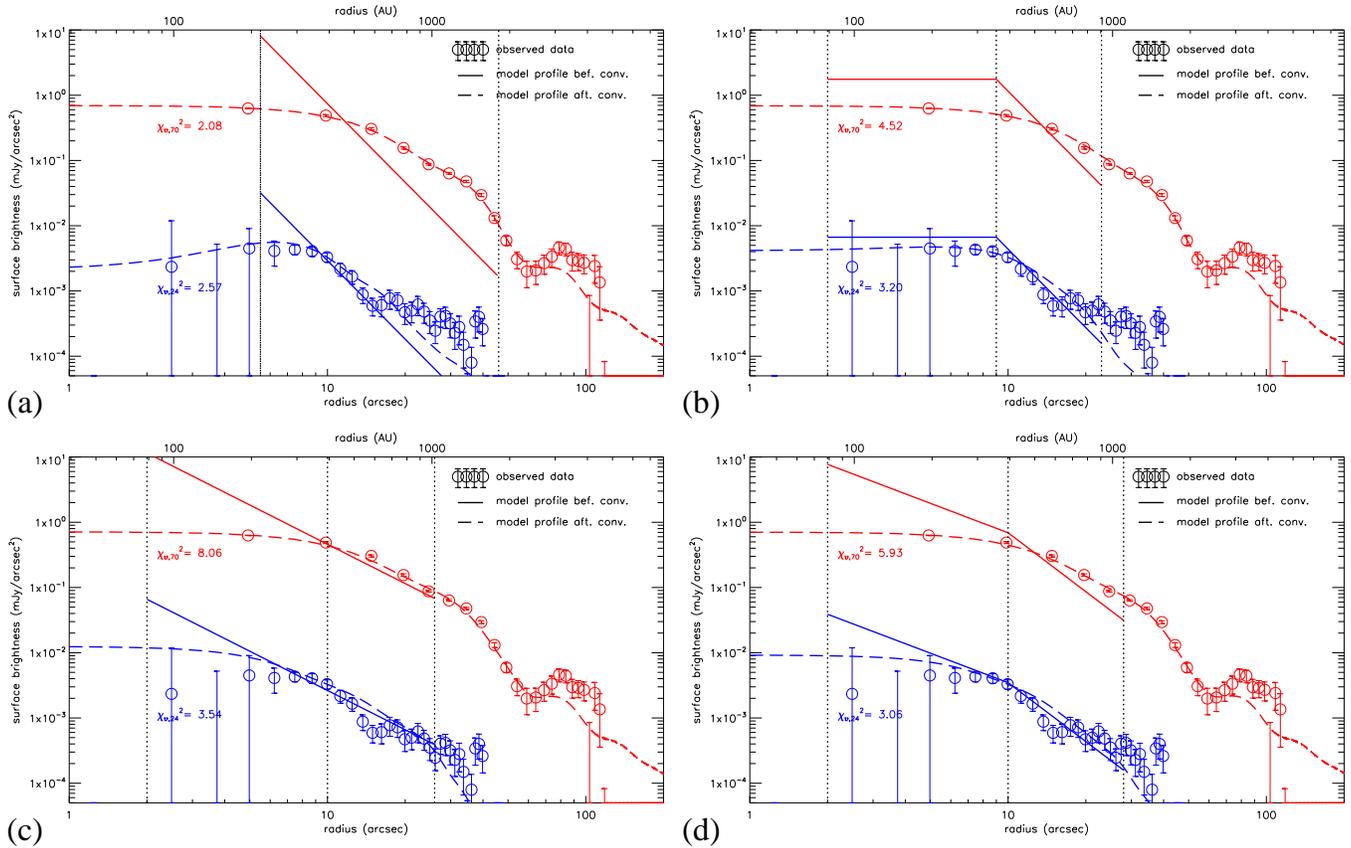}
\caption{Model and observed surface brightness distributions at 70 \um
  (red) and 24 \um (blue). Solid and dashed-lines are the model before
and after PSF convolution, respectively. For details about the model
surface brightness parameters see \S
\ref{analysis:surface_brightness_distribution}. In all cases, $S(r)=$
0 for $r\le$2\arcsec. (a) is for a large, empty inner hole: $r_h =
r_{br}=$5\farcs5, $S(r)\sim r^{-4}$, $r_{out}=$46\arcsec; (b) a flat
inner distribution: $r_{h}=$2\arcsec, $r_{br}=$9\arcsec, $S(r)\sim
r^{-4}$, $r_{out}=$23\arcsec; (c) a single power law:
$r_{h}=$2\arcsec, $r_{br}=$10\arcsec, $S(r)\sim r^{-2}$,
$r_{out}=$26\arcsec; (d) two power laws: $r_{h}=$2\arcsec, $S_1(r)\sim
r^{-2}$, $r_{br}=$10\arcsec, $S_2(r)\sim r^{-3}$ ,
$r_{out}=$28\arcsec. The best fit (indicated by lowest $\chi^2_{\nu}$) is
case (a) with $r_{out}=$46\arcsec.}   
\end{figure*}

We have tried four different cases for the model surface brightness
distribution: (1) a larger inner hole with $\alpha$=4, (2) a flat
distribution between $r_{h}$ and $r_{br}$ but with $\alpha$=4 outward,
(3) a single powerlaw with $\alpha$=2, and (4) two powerlaws with
indices of $\alpha$=2 and $\alpha$=3. In all cases, we varied $r_{br}$
and $r_{out}$ of the surface brightness distribution (two free
parameters) and computed the reduced $\chi^2_{\nu}$ value to determine the
best fit. The reduced $\chi^2_{\nu}$ value was computed for data points
between 5\arcsec--40\arcsec, and between 5\arcsec--75\arcsec~at 24 and
70 \mm, respectively. Note that the bright hump seen in the 70 \um
profile between 75\arcsec~and 90\arcsec~is where the second bright
Airy ring is located (see Fig.~\ref{radprof70}). The match between an
observed PSF derived from the calibration stars and theoretical PSF is
poor in this range; therefore, we disregard the 70 \um data points
outside 75\arcsec~in the modeling. The results are shown in Figure
\ref{radprof_fits}.

\begin{figure}
\figurenum{8}
\label{radius_fr}
\plotone{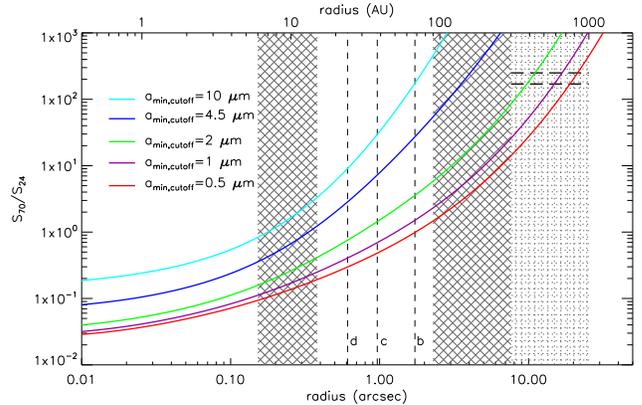}
\caption{Radial-dependent surface brightness flux ratio
  ($S_{70}/S_{24}$) in the HR 8799 system for various sizes of
  minimum grain size cutoffs. As in Fig.~\ref{radius_td}, the projected distances
  of the three planets are marked as dashed lines. The three disk
  components are marked in different filled patterns. The range of the
  flux ratios inferred from the models in \S
  \ref{analysis:surface_brightness_distribution} are marked as the long-dashed
  lines near the upper right corner.} 
\end{figure}

Due to {\it Spitzer}'s large beam sizes in the far-infrared, the
assumed surface brightness behavior and its number of free parameters
needed to describe the model, there is no unique fit for the observed
profiles. The best-fit outer radii range from 23\arcsec~to 46\arcsec,
consistent with the previous finding of emission from outside of the
submillimeter planetesimal ring ($<$10\arcsec). From the model surface
brightness profiles, we can also derive the surface brightness ratio
($S_{70}/S_{24}$) in the extended disk. The ratios range from
$\sim$170 to $\sim$250, which can be used to set constraints on the
grain properties since the density-dependent term cancels out in the
ratio. Figure \ref{radius_fr} shows the expected ratio for a size
distribution of $a^{-3.5}$ with a maximum size cutoff of $\sim$10 \um
and various minimum grain size cutoffs ($a_{min,cutoff}$) using
astronomical silicate grains. Based on the range of derived
$S_{70}/S_{24}$ ratios (represented by the horizontal dashed line in
figure \ref{radius_fr}), the acceptable range of $a_{min,cutoff}$ is
$\lesssim$2 \um (=$a_{bl}$). Outside the main planetesimal disk
($r>$10\arcsec), only grains with minimum sizes $\lesssim$ the
radiation blowout size have flux ratios in the observed range. This
supports the idea that the disk outside of the main planetesimal disk
consists of small grains being pushed outward by radiation pressure,
i.e., the HR 8799 disk has an extended halo similar to that seen
around the Vega disk \citep{su05}.

\begin{figure*}
\figurenum{9}
\label{models}
\plotone{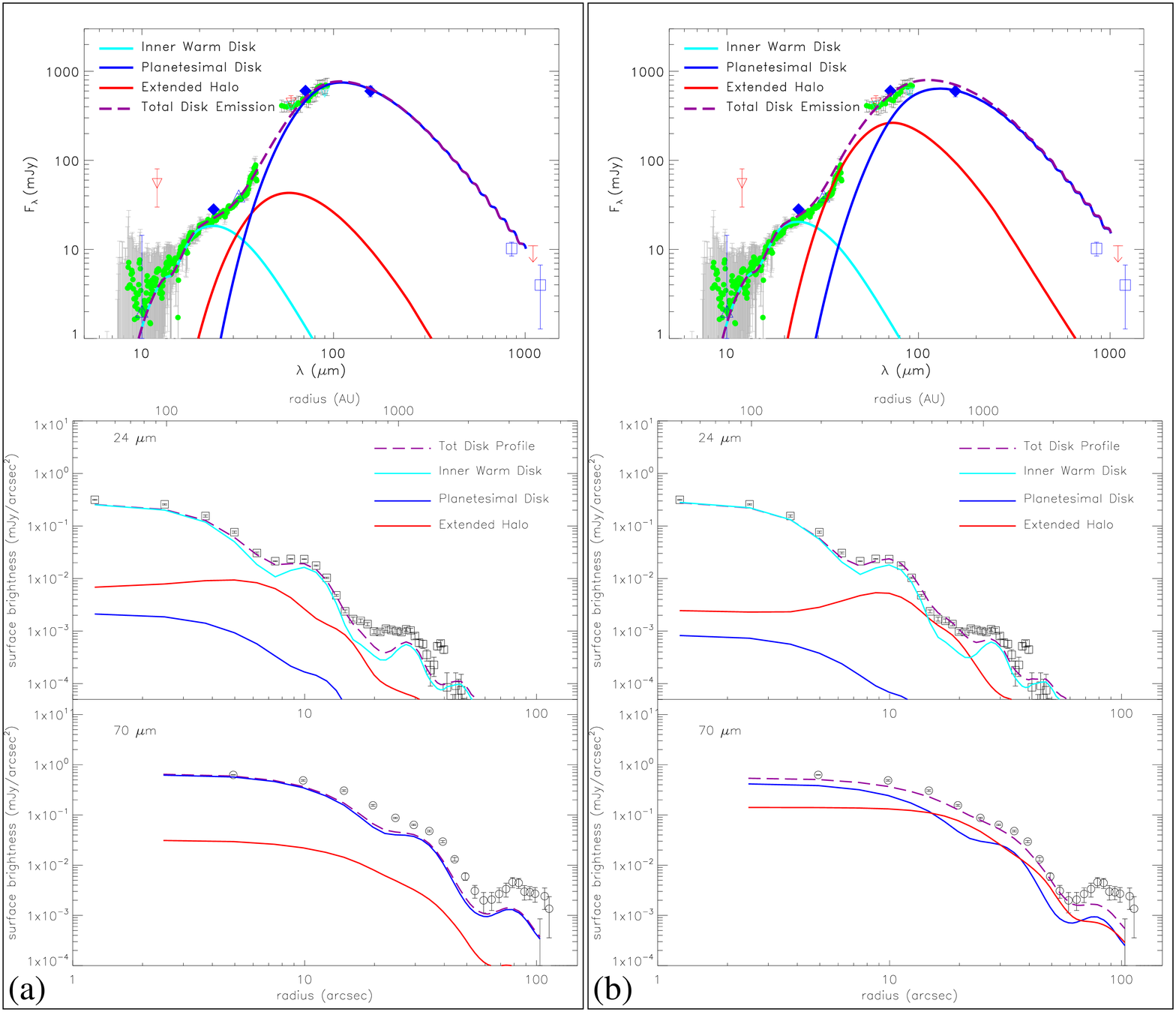}
\caption{Model SED (top panel) and surface brightness profiles (bottom
  panel) compared to the data. Two sets of models are shown having
  the same inner warm disk component ($R_{in,warm}$=6 AU,
  $R_{out,warm}$=15 AU, $a_{min}$=1.5 \mm, $a_{max}$=4.5 \um with total
  $M_d$ of 1.1$\times$10$^{-6}M_{\earth}$) and (a) a narrow planetesimal
  disk model ($R_{in,pb}$=90 AU, $R_{out,pb}$=150 AU, $a_{min}$=10 \um and
  $a_{max}$=1000 \um with $M_d$=5.5$\times$10$^{-2}M_{\earth}$) plus 
  a halo ($R_{in,halo}$=150 AU, $R_{out,halo}$=1000 AU,
  $a_{min}$=1 \um and $a_{max}$=1 \um with
  $M_d$=1.2$\times$10$^{-3}M_{\earth}$), or (b) a wide planetesimal disk
  model ($R_{in,pb}$=90 AU, $R_{out,pb}$=300 AU, $a_{min}$=10 \um and
  $a_{max}$=1000 \um with $M_d$=1.2$\times$10$^{-1}M_{\earth}$) plus
  a halo ($R_{in,halo}$=300 AU, $R_{out,halo}$=1000 AU,
  $a_{min}$=1 \um and $a_{max}$=10 \um with 
  $M_d$=1.9$\times$10$^{-2}M_{\earth}$). The narrow ring planetesimal
  disk model gives a satisfactory fit to the SED, but fails to match
  the disk radial profiles in contrast to the fit of a wide
  planetesimal disk shown in (b).  
  The model submillimeter
  flux appears to be too high compared to the observed data. However,
  because of the source extension, the ``observed'' model flux
  actually agrees with the measurements that were calibrated assuming an
  unresolved source. For details see \S \ref{dis_cold_comp}.}
\end{figure*}

\subsection{Specific Disk Models}
\label{analysis:thermal_model}

From the preceding section, we can expand on the description of the
three disk components summarized at the end of \S 3.2: (1) the warm
belt inside the innermost planet~d has a temperature of $\sim$150 K
with parameters dubbed as ``$warm$''; (2) the main planetesimal disk
outside the outermost planet~b has a temperature of $\sim$45 K with
parameters dubbed as ``$pb$'' (parent bodies); and (3) the outermost
component is small grains strongly affected by photon pressure with
parameters dubbed as ``$halo$''. The basic equations for computing the SED
and surface brightness of an optically-thin debris disk can be found
in \citet{backman93}, \citet{wolf03}, and \citet{su05}. 
To model these structures, we use grain
properties for astronomical silicates with a density of 2.5
g~cm$^{-3}$ and a $n(a)\sim a^{-3.5}$ size distribution appropriate
for a steady-state collisional cascade \citep{dohnanyi69}. Using the SED of the excess
and the resolved images as constraints, we fit the minimum and
maximum grain sizes ($a_{min}$ and $a_{max}$) in the size distribution
and the inner and outer radii of each of the disk components ($R_{in}$
and $R_{out}$) by assuming a constant surface density ($\Sigma(r)\sim
r^0$) disk for the strongly bound disk and $\Sigma(r)\sim r^{-1}$ for the
weakly-bound or unbound disk (i.e., the grains affected by photon pressure).

From its characteristic temperature ($\sim$150 K) and Figure
\ref{radius_td}, the inner radius of the warm component can be as
close as $\sim$6 AU if the grains have a size of $\sim$10
\mm. However, the spectral shape of the 10 to 20 \um excess emission
is better fit with grains of $\sim$\um sizes. In addition, the
spectral shape suggests that sub-micron grains are absent, since their
emission spectrum would result in higher contrast features at 10 and
20 \mm. We fit the inner warm component as a bound disk with
parameters used in Table \ref{model_param}.  The dust mass only
accounts for the small grains that dominate the mid-IR emission. We
lack the constraints, such as the flux of this warm component at
longer wavelengths, that would be required to estimate the mass of the
population of large grains and parent bodies associated with the warm
emission.

It is less straightforward to fit the other two disk components, due
to the lack of strong constraints on the size of the planetesimal disk.
We can, however, use the following information: (1) we know roughly
how much flux comes 
from the unresolved component (presumably the planetesimal disk) at 70
\mm; (2) the minimum inner radius of the planetesimal disk is $\sim$90
AU, from the characteristic temperature of the cold component in the
SED; and (3) large grains (at least a few hundred microns) are
required due to the collisional cascade nature of debris disks. The
model parameters not only have to provide satisfactory fits to the
global SED, but the model images after convolution with the beam sizes
have to match the observed images as well. Our modeling strategy is
similar to the one applied to the $\gamma$ Oph disk \citep{su08}; we
first tried many combinations of parameters that can provide good fits
to the SED, then refined the parameters by matching the observed
radial profiles. We started with constraining the parameters for the
inner component using data shortward of 30 \um since they are most
sensitive to this component, then used the constraints listed above for
the parameters in the other two disk components.

For the planetesimal disk, we fixed $a_{max}$ and $R_{in,pb}$ at 1000
\um and 90 AU, respectively. Additionally, we constrained $a_{min}$ to
be greater than $a_{bl}$ ($\sim$2 \mm). For the extended halo disk,
$R_{out,halo}$ was fixed at 1000 AU; $R_{in,halo}$ and $R_{out,pb}$
were required to be equal (there is no gap between the planetesimal
and halo disks). Initially, we fit the planetesimal disk with a narrow
ring ($\Delta R<R$) so the actual density distribution has less effect
in the output SED.  A difficulty in this narrow ring model is that
grains smaller than $a_{bl}$ have relatively high thermal equilibrium
temperatures at the location where the extended disk starts, and the
resultant emission from the inner edge of the halo component is too
strong in the 25--35 \um range. The only way to make the narrow ring
model work is to have far less contribution from the extended disk as
shown in the SED of Figure \ref{models}a (the top panel). However, the
resulting model yields insufficient flux from the extended component,
and therefore, also shows a disk at 70 \um that is too small, as shown
in the model radial profile (Fig.~\ref{models}a).

To satisfy all the constraints listed above simultaneously, the outer
radius of the planetesimal disk has to extend to $\sim$300 AU. Figure
\ref{models}b shows one of the best-fit models in terms of SED and
radial profiles. In this model, the planetesimal disk extends from
$R_{in,pb}$=90 AU to $R_{out,pb}$=300 AU and consists of grains of
$a_{min}$=10 \um to $a_{max}$=1000 \um with a total dust mass ($M_d$)
of 1.2$\times$10$^{-1}M_{\earth}$. The halo disk extends from
$R_{in,halo}$=300 AU to $R_{out,halo}$=1000 AU and consists of grains of
$a_{min}$=1 \um to $a_{max}$=10 \um with a total
$M_d$=1.9$\times$10$^{-2}M_{\earth}$. Note that the reduced
$\chi^2_{\nu}$ values in the SED fitting are similar between the
narrow-ring and broad-disk models; however, the model flux ratio at 70
\um between the extended component (halo) and the unresolved core
(planetesimal disk) is too low compared to the observed ratio
($\sim$1). This mis-match is best shown in the model surface 
brightness profile at 70 \um (bottom panel of Fig.~\ref{models}a),
where the narrow-ring model does not fit the 70 \um profile at all.

The parameters used and derived in our thermal model for the 
disk components are listed in Table \ref{model_param}. 
Because of the ambiguous nature of SED modeling with multiple
components and low-resolution images, these model parameters are not
unique but were built from the least number of components with simple
assumptions that produced a good match to all the available data. One
should not take the model grain sizes too literally since they
depend greatly on the grain properties used. However, the total dust
mass in each of the components remains similar among all of the
acceptable models.

\begin{deluxetable}{cccc}
\tablewidth{0pc}
\tablecaption{Parameters in the Preferred Model\label{model_param}}
\tablehead{ 
\colhead{Parameters} & \colhead{Inner Warm Disk} & \colhead{Planetesimal Disk} &
\colhead{Halo} \\
\colhead{} & \colhead{($warm$)} & \colhead{($pb$)} & \colhead{($halo$)}  
}
\startdata
$\Sigma(r)$  &  $\sim r^0$  &  $\sim r^0$  &  $\sim r^{-1}$  \\ 
$R_{in}$ (AU)    &  6   &  90  &  300    \\
$R_{out}$ (AU)   & 15   & 300  & 1000    \\
$a_{min}$ (\mm)  & 1.5  &  10  & 1       \\
$a_{max}$ (\mm)  & 4.5  & 1000 & 10      \\
$M_d$ ($M_{\earth}$)& 1.1$\times$10$^{-6}$ & 1.2$\times$10$^{-1}$ &
1.9$\times$10$^{-2}$ \\ 
$f_d=L_{IR}/L_{\ast}$ &
2.2$\times$10$^{-5}$&7.5$\times$10$^{-5}$&1.0$\times$10$^{-4}$ 
\enddata 
\end{deluxetable}

\section{Discussion}

\subsection{Debris Disk Structure}

The architecture of the HR 8799 planetary system is complex: 
a warm inner asteroid-belt analog located at a radius of 6--15 AU; a
cold Kuiper-belt-analog planetesimal disk located from $\sim$90 AU up
to $\sim$300 AU; three massive planets orbiting between the two disk
components; and a prominent halo of small grains extending up to
$\sim$1000 AU. Similar components (except
for imaged planets) have been found in a number of other debris disk
systems, indicating some underlying order within the diversity in
debris disk structures.

\subsubsection{Inner Warm Component} 

Assuming the grains in the inner belt are astronomical silicates, we
can rule out grains of sub-micron size, due to the 
absence of strong silicate features at 10 and 20 \mm. The shape of the
IRS spectrum strongly favors micron sizes for the grains that dominate
the emission. If we further assume that all the grains in this
component are bound (a$_{min}$ = a$_{bl}$), then the outer radius of
this warm component can be as small as $\sim$10 AU based on
temperature arguments (see the 2 \um curve in
Fig.~\ref{radius_td}). We also find that the 
maximum inner radius of this component is around $\sim$6 AU, again
from temperature arguments. There are few, if any, grains inside this
radius.  An upper limit on the dust mass inside this inner warm
component can be estimated based on the observed IRS excess
spectrum. A flux density of 1 mJy at 10 \um corresponds to a dust mass of
2.3$\times$10$^{-7}M_{\earth}$ assuming astronomical silicates of
$a$=10 \um at 3 AU (i.e., T$_d\sim$200 K) and a grain density of 2.5 g
cm$^{-3}$. This inner hole can be maintained by ice sublimation if the
grains are icy \citep{jura98}; a process that will occur at a dust
temperature of $\sim$150 K. Alternatively, the lack of grains inside
$\sim$6 AU suggests either that there is another (unseen) interior
planet maintaining a dust-free zone, or that the growth of the
protoplanets in the terrestrial zone (3--20 AU for an A-type star) at
an early stage initiated a collisional cascade that moved outward and
created this inner hole \citep{kenyon04}.  Since we lack constraints
such as a measurement of the long wavelength emission from this inner
component at higher spatial resolution, we cannot set
limits on $a_{max}$. Because the 
dust in this component is warm, very little mass is required, only
about 1.1$\times$10$^{-6}M_{\earth}$ for dust with an
infrared fractional luminosity ($f_d$) of 2.2$\times$10$^{-5}$. These
values are rough lower limits because it is difficult to place tighter
constraints without knowing additional parameters such as $a_{max}$.

We also fit the inner warm component with amorphous carbon grains
(density of 1.85 g/cm$^3$, \citealt{zubko96}). As expected from the
discussion in \S~\ref{sec_models}, the model results for the placement
of the disk are not very sensitive to the assumed grain
composition ($R_{in,warm}\sim 6$ AU and $R_{out,warm}\sim$15 AU), and
we can also rule out a significant amount of sub-micron carbonaceous
grains in this inner component because the equilibrium temperature of
150 K places them right at the location of planet c. A similar result of
low sensitivity to grain properties is
also found in \citet{su05}, who analyzed the behavior of the
Vega system for grains of silicate, carbonaceous, and mixed
composition.

A similar inner warm component (T$_d \sim$150 K) that dominates the
disk emission at the IRS and MIPS 24 \um wavelengths is seen
relatively commonly among resolved debris systems. For example, HR
4796A, Fomalhault, and $\epsilon$ Eri disks all have such a component
\citep{wahhaj05,stapelfeldt04,backman09}. In fact, all of their excess
SEDs look very similar: a steep inflection between the IRS spectrum
and MIPS-SED mode data, indicating a relatively sharp inward-facing
edge to the structure dominating the far-infrared emission. This
feature may arise from the action of massive planets, as appears to be
the case for Fomalhaut \citep{kalas08,quillen06,chiang09}. HD~32297
and $\eta$ Tel are other systems that show similar characteristics
revealed by recent ground-based imaging \citep{moerchen07,fitzgerald07,
smith09}. In addition, a recent IRS spectroscopic study of unresolved
debris disks shows that warm components are very common ($\sim$50\%)
among A-type excess stars \citep{morales09}. In the case of HR 8799,
there are strong indications that this warm component is physically
separated from the cold, outer component and that there is little dust
in the intervening region.

\subsubsection{Outer Cold Planetesimal Disk} 
\label{dis_cold_comp} 

Our best-fit model appears to indicate too much
flux in the submillimeter (Fig.~\ref{models}) for the cold
planetesimal disk. However,
the model 850 \um disk has a FWHM of 17\farcs8 after being convolved
with the nominal 850 \um beam size of 15\arcsec. This slight extension
would not be resolved by the 850 \um observation because the data were
taken in photometry mode and no mapping was done. The flux density
reported in \citet{williams06} was 
calibrated assuming a point source. Therefore, the correct way to
compare the model and observed fluxes is to integrate the total
flux in a 15\arcsec~diameter aperture on the model image. The 850 \um
model flux within a diameter of 15\arcsec~is 9.33 mJy, consistent with
the SCUBA measurement of 10.3$\pm$1.8 mJy. Similarly, the 1.2 mm flux
is also under-estimated for the source extension since it was obtained
using the IRAM 30 m single dish telescope (FWHM=9\farcs5).

The dust mass in the planetesimal disk that is derived
from our three-component model is 0.12 $M_{\earth}$, consistent with
the mass derived from the 850 \um observation (0.1 $M_{\earth}$,
\citealt{williams06}). This is only twice as large as the mass
required by the narrow ring model, even though the disk is much larger.
The dust mass in this cold component is similar to values derived for
disks in the 12 Myr-old $\beta$ Pic moving group ($\beta$ Pic and
HD~15115, \citealt{holland98, williams06}). The amount of the excess
emission in the HR 8799 disk ($R_{24}$=1.5 and $R_{70}$=94, compared
to the stellar photosphere) is comparable with other debris disks
around A-stars of similar age (20--160 Myr) \citep{su06}, at least
within statistical expectations. The infrared fractional luminosity
for this component is 7.5$\times$10$^{-5}$, suggesting a
collision-dominated disk. 

Since the cold component is the birth place for the second-generation
debris from collisional cascades, the presence of large grains is
required. The inner edge of the disk, although
not directly resolved by imaging, has a minimum radius of $\sim$90 AU
from temperature arguments. We lack high-spatial-resolution
submillimeter data to put strong constraints on the outer edge of the
planetesimal disk. The outer radius of $\sim$300 AU in our model is 
based on the fewest and most straightforward assumptions. 

\subsubsection{Halo}

Finally, we consider the halo component of the system. The general
appearance and extent of this component are similar to the halo around
the Vega system. In that case, the {\it Spitzer} observations provided
sufficient angular resolution to fit the surface brightness profile
and show that it is consistent with expectations for grains being blown out
by radiation pressure. While the details of this process are complex,
it is generally parametrized by $\beta$, the ratio of radiation pressure
force to gravitational force. When collisions produce grains with
$\beta$ $>$ 1, they will be blown out of the system 
quickly. When the collisional cascade produces grains over a narrow
range of $\beta$ below unity, they will tend to go into elliptical
orbits. The behavior of these barely-bound grains has been suggested
to explain the broken power-laws seen in the scattered-light disk
surface brightness profile around AU Mic \citep{strubbe06} and $\beta$
Pic \citep{augereau01}.  However, these highly-elliptical orbits will
continue to cross the parent body 
zone and eventually it is likely that these grains will collide again,
be broken into smaller ones with $\beta > 1$ and be blown out of the
system. Thus, the time scale for mass loss from the system depends on
the details of the collisional fragmentation of the grains. 

However,
qualitatively, the halo around HR 8799 resembles the behavior of Vega
closely and differs substantially from the observations of most other
A-stars, indicating that both systems are currently undergoing an
enhanced level of dynamical stirring in their outer planetesimal
disks that leads to an elevated production rate for small grains with
$\beta$ near unity. These grains will eventually be ground down to
sizes that participate in a general outflow from the system.  We
therefore model the halo by making assumptions similar to those used for
the Vega outflow \citep{su05}.

To fit the Vega outflow, grains extending up to about
five times the nominal blowout size were required. This apparent
discrepancy probably arises because, for simplicity, the models are
based on solid, spherical grains, whereas real grains are likely to be
non-spherical and complex in structure with a substantial volume of
voids (e.g., \citealt{dominik07} and references therein).  The
influence of photon pressure will be much larger for such grains than
for solid spherical ones. We use $a_{max} = 10$ \um in
the extended halo component,
about five times larger than the blowout size for HR 8799, and other
parameters of the model are similar to those used for Vega
\citep{su05}. The estimated dust mass in the halo is 
relatively large, 1.9$\times 10^{-2} M_{\earth}$ that is $\sim$6 times
greater than the mass in the Vega outflow. The terminal radial
velocity is 2--4 km~s$^{-1}$ for grains having $\beta \sim 1$
and that are being produced in the planetesimal disk, 100--300 AU from
a 1.5 M$_{\sun}$ star \citep{amaya05}. For a disk
radius of 1000 AU, the residence time for such particles is $\sim$
900--2400 yr. For a static collisionally dominated disk
\citep{wyatt07}, the expected ratio of mass in blown out grains to
that of grains in the bound disk is $\sim$1.2\% in the HR 8799 system (i.e.,
a 1.5 M$_\odot$ star with a fractional disk luminosity of 8$\times
10^{-5}$ and $r \sim$ 100 AU). The observed ratio for HR 8799 is 15
times higher, qualitatively similar to the Vega system, where the mass
in blown out grains is about 30 times the predicted static level. Both
systems have much more prominent halos than systems such as Fomalhaut
\citep{stapelfeldt04} and $\beta$ Leo (Stock et al. in prep.). This
comparison indicates that there is an elevated 
level of dynamical activity in the bound disks of Vega and HR 8799
that is enhancing the production of small grains.

Although we have proceeded by analogy with Vega, future work should
address the role of weakly bound grains in more detail. The conclusion
that the halo indicates more dynamical activity than in A stars
without such 
halos should be robust, but more complete modeling should draw out
other aspects of the HR 8799 system. 

\subsection{Interaction of Planets and Debris}

Based on the ``resonance overlap'' condition in the planar circular
restricted three body problem, the width of the chaotic region,
$\Delta a$, in the vicinity of a planet with mass of $M_p$ and orbital
semi-major of $a_p$ orbiting a star with mass of $M_{\ast}$ is given
by \citet{malhotra98}, 
\begin{equation}
\Delta a \simeq 1.4 a_p (M_p/M_{\ast})^{(2/7)}.
\end{equation}
Adopting the nominal parameters from \citet{marois08}, stellar mass
1.5 $M_{\sun}$, planet masses 7 $M_J$ and 10 $M_J$, and semi-major
axes 24 or 68 AU for planets d and b, respectively, the width of the
unstable zone near each planet is is $\sim$8 and $\sim$23 AU, respectively. 
Therefore, the locations of the outer edge of
the inner dust belt ($\sim$15 AU) and the inner edge of the outer belt
($\sim$90 AU) are both consistent with gravitational
sculpting by the innermost known planet d and the outermost known
planet b, respectively. 
The nearly dust-free zone interior to $\sim$6
AU may indicate the presence of additional unseen inner planets. If
so, then this leads to an estimate of the timescale of inner planet
formation of $<$20--160 Myr given the age of HR 8799. 
This is in good agreement with theoretical calculations for
terrestrial planet accretion \citep{wetherill92} as well as isotopic
constraints on the age of the Earth \citep{podosek00}.

The existence of the halo indicates that the outer planetesimal
disk is heavily stirred due either to: (1) the planets not being in a stable
configuration \citep{fabrycky08,reidemeister09,goz09};
(2) planet migration and orbital resonance crossings
that are causing extreme excitation in the planetesimal disk, as in
models for  
the Kuiper Belt  \citep{malhotra95} and for the Late Heavy
Bombardment \citep{gomes05}; or 
(3) processes in the planetesimal disk itself, such as ongoing
formation of (unseen) ice giants or planetary embryos that stir the
disk \citep{kenyon08}.  The first two possibilities relate the
dynamical activity in the planetesimal disk to issues associated with
the stability of the configuration of the three known planets, while
the third hypothesis would be independent of the planets.  Although
no specific conclusions can be drawn, the outcomes of any of these
scenarios are a heavily stirred planetesimal disk that produces an era
of enhanced collisional activity. The resulting collisional cascades
and avalanches will feed an outflow of small grains \citep{grigorieva07}.

\begin{figure}
\figurenum{10}
\label{a160_co32}
\plotone{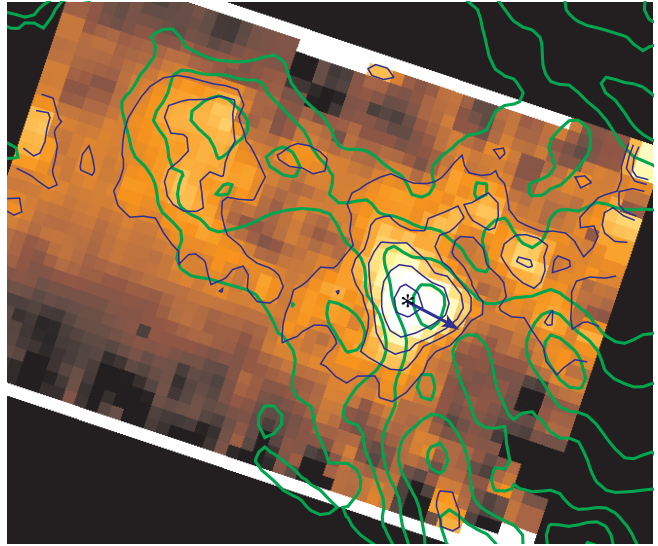}
\caption{The CO J=3-2 contours (green) overlaid on the 160 \um data
  (pseudo-color with blue contours). The field of view is
  7\arcmin$\times$5.9\arcmin~with N up and E toward the left. The
  contours of the 160 \um data range from 1-, 2-, 3-, 5- and
  8-$\sigma$ detection level, while the green CO contours show 5 linear
  levels of integrated intensity from 0.94 (2-$\sigma$) to 2.7
  (7-$\sigma$) K km~s$^{-1}$. The
  stellar position is marked as a star-shaped symbol and the blue
  arrow shows the proper motion direction of HR 8799 with a length indicating 40\arcsec.} 
\end{figure}

\subsection{The Background Cloud near HR 8799 and the $\lambda$-Bootis
  Phenomenon} 
\label{bkg_cirrus}

The cirrus seen at 160 \um resembles some of the large-scale structure
in the CO (3-2) cloud detected by \citet{williams06}. Figure
\ref{a160_co32} shows a new, 7\arcmin$\times$7\arcmin~CO (3-2) map at
15\arcsec~resolution around HR 8799, taken with the HARP heterodyne
array on the JCMT, overlaid on the {\it Spitzer} 160 \um map.  The CO
emission is integrated in the local standard of rest (LSR) velocity
from -4 to -6 km s$^{-1}$. The radial velocity of the star,
-12.4$\pm$0.5 km s$^{-1}$, from \citealt{moor06} is in the
heliocentric velocity frame, however, and 
converts to -4.6 km s$^{-1}$ LSR.  Thus, contrary to the conclusion in
\citet{williams06}, we cannot say that the cloud is not associated
with HR 8799 and cannot place any limits on the gas content of the disk
from these observations. This is also further confirmed by the new
IRAM 30-m observation that found strong CO (2-1) line emission due to
the cloud at the heliocentric radial velocity of HR 8799 (Hily-Blant,
Kastner, Forveille \& Zuckerman, in prep.)  

This new JCMT CO (3-2) map is larger and of higher quality than the map presented
in \citet{williams06} and shows the extended, low level molecular
emission in more detail. It appears to be associated with the extended
arc of high latitude clouds, MBM~53-55, detected in CO (1-0) by
\citet{magnani00}.  The distance of these objects and their relation
to one another is not well known.  If the association with the star is
physical then, at 40~pc, this would be the closest known molecular
cloud. 

Whether physically associated or not, it is likely that the low flux
levels (1-3$\sigma$) of the 160 \um emission come from the molecular
cloud.  The asymmetry in the outer part of the observed source at this
wavelength may also be due to cloud contamination.  Nevertheless, the
detected source (with the peak flux $>$ 8-$\sigma$) is coincident with
the expected stellar position, which is offset ($\sim$20\arcsec) from
the central emission of the nearby clump in the cloud. The cloud
therefore contributes only a very small amount of flux that cannot be
completely removed from our aperture photometry. At 70 \mm, there is
no resemblance to the large-scale background structure in the data nor
an asymmetry seen in the source morphology; therefore, we can rule out
any significant contribution from this extended background cloud to
the 70 \um disk emission.

The arrow in Figure \ref{a160_co32} shows the direction of proper
motion of HR~8799 ($\delta$RA = 107.93 mas yr$^{-1}$ and $\delta$Dec
= --49.63 mas yr$^{-1}$; \citealt{vanleeuwen07}). Given the
coincidence in space and velocity, it is plausible that accretion from
the cloud might explain the abundance anomalies leading to the
$\lambda$ Bootis designation \citep{kamp02}. Using equation (16) from
\citet{martinez09} for Bondi-Hoyle accretion implies a mass accretion
rate of $\sim 3 \times 10^{-14} M_{\sun} yr^{-1}$ with an assumed
density of 100 cm$^{-3}$, just above the threshold to produce
$\lambda$ Bootis behavior \citep{turcotte93}. However, this
calculation assumes a fluid model for the ISM, which is not strictly
applicable at the expected density \citep{martinez09}. Therefore, the
accretion rates may be significantly lower than indicated
\citep{alcock80}.  The accretion hypothesis rests on the untested
assumption that the rate can be increased through such means as small
scale high-density structures in the ISM and/or increases in density
in the wake of the star.

\section{Conclusions}

We have obtained deep infrared images and spectra of the debris system
around HR 8799. To our surprise at the distance of $\sim$40 pc, the
HR~8799 disk is clearly resolved at both 24 and 70 \mm; the outer
boundary of the disk can be traced further than 1000 AU.  
From general, model-independent considerations, the
star appears to have one zone of warm dust within the orbit of its
innermost known planet, another broad zone of cold dust outside the orbit of
its outermost known planet, and an extended halo of small grains from this
outer zone. 

We have constructed detailed models of the system, guided by models
of other debris systems as well as the observations of HR
8799. Although the models are not unique, because we draw on
experience in modeling many debris systems, they should give a
reasonably accurate picture of conditions in the HR 8799 system. The
models indicate that the inner zone lies between 6 and 15 AU with a
total dust mass of 1.1$\times$10$^{-6}M_{\earth}$ and the outer zone
is between 90 and 300 AU with a dust mass of
1.2$\times$10$^{-1}M_{\earth}$. There is little dust inside the inner
zone ($r\lesssim$6 AU) and in between the inner and outer cold zones
where the three planets reside. The halo extends to at least 1000
AU and has a dust mass of 1.9$\times$10$^{-2}M_{\earth}$, about 15
times the level expected for a static, quiescent debris system. The
implied high level of dynamical activity in the HR 8799 debris system
may be related to perturbations imposed by its system of three
massive planets or may be signaling ongoing planet formation in the
outer parts of this system.

HR 8799 is at an age similar to that of the solar system when the 
terrestrial planets formed, along the way to settling into the final
configuration. The planetary system of HR 8799 (both planets and
debris disk) provide an intriguing snapshot of processes occurring at
this stage and may help us understand the formation and evolution of
our own planetary system.

\acknowledgments
This work is based on observations made with the {\it Spitzer Space Telescope},
which is operated by the Jet Propulsion Laboratory, California
Institute of Technology. Support for this work was provided by NASA
through contract 1255094 and 1256424 issued by JPL/Caltech to the University of
Arizona. We thank Chian-Chou Chen for assistance with the JCMT CO 
data, David Wilner and Eric Mamajek for discussion of the
background cloud, and Ben Zuckerman for pointing out the error in the
velocity comparison of the star and the cloud. KS also thanks Glenn
Schneider and Mike Meyer for the useful discussion. This research has
made use of the SIMBAD database, operated at CDS, Strasbourg, France.


\begin{thebibliography}{}


\bibitem[Alcock \& Illarionov(1980)]{alcock80} Alcock, C., \& Illarionov, A.\ 1980, \apj, 235, 541 

\bibitem[Augereau et al.(2001)]{augereau01} Augereau, J.~C., Nelson, R.~P., Lagrange, A.~M., Papaloizou, J.~C.~B., \& Mouillet, D.\ 2001, \aap, 370, 447 


\bibitem[Backman \& Paresce(1993)]{backman93} Backman, D.~E., \& Paresce, F.\ 1993, Protostars and Planets III, 1253 
 
\bibitem[Backman et al.(2009)]{backman09} Backman, D., et al.\ 2009, \apj, 690, 1522 

\bibitem[Beichman et al.(2005)]{beichman05}Beichman, C. A. et al. 2005, \apj, 626, 1061

\bibitem[Bondi \& Hoyle(1944)]{bondi44} Bondi, H., \& Hoyle, F.\ 1944, \mnras, 104, 273 


\bibitem[Castelli \& Kurucz(2003)]{castelli03} Castelli, F., \&
Kurucz, R.~L.\ 2003, IAU Symposium, 210, 20P

\bibitem[Chen et al.(2006)]{chen06} Chen, C.~H., et al.\ 2006, 
\apjs, 166, 351 

\bibitem[Chen et al.(2009)]{chen09} Chen, C.~H., Sheehan, P., Watson, D.~M., Manoj Puravankara, P., \& Najita, J.~R.\ 2009, arXiv:0906.3744 


\bibitem[Chiang et al.(2009)]{chiang09} Chiang, E., Kite, E., Kalas,
  P., Graham, J.~R., \& Clampin, M.\ 2009, \apj, 693, 734  

\bibitem[Cowley et al.(1969)]{cowley69} Cowley, A., Cowley, C., 
Jaschek, M., \& Jaschek, C.\ 1969, \aj, 74, 375 

\bibitem[Dohnanyi(1969)]{dohnanyi69} Dohnanyi, J.~W.\ 1969, \jgr, 74,
2531

\bibitem[Dominik et al.(2007)]{dominik07} Dominik, C., Blum, J., Cuzzi, J.~N., \& Wurm, G.\ 2007, Protostars and Planets V, 783 


\bibitem[Engelbracht et al.(2007)]{engelbracht07} Engelbracht, C.~W.,
et al.\ 2007, PASP,119, 994


\bibitem[Fabrycky \& Murray-Clay(2008)]{fabrycky08} Fabrycky, D.~C., \&
Murray-Clay, R.~A.\ 2008, arXiv:0812.0011 

\bibitem[Fitzgerald et al.(2007)]{fitzgerald07} Fitzgerald, M.~P., 
Kalas, P.~G., \& Graham, J.~R.\ 2007, \apj, 670, 557 


\bibitem[Go\'zdziewski \& Migaszewski (2009)]{goz09}Go\'zdziewski,
  K. \& Migaszewski, C. 2009, astro-ph/0904.4106

\bibitem[Gomes et al.(2005)]{gomes05} Gomes, R., Levison, H.~F.,
  Tsiganis, K., \& Morbidelli, A.\ 2005, \nat, 435, 466  

\bibitem[Gordon et al.(2005)]{gordon05} Gordon, K.~D., et al.\ 2005,
\pasp, 117, 503

\bibitem[Gordon et al.(2007)]{gordon07} Gordon, K.~D., et al.\ 2007, PASP, 119, 1019

\bibitem[Gray \& Kaye(1999)]{gray99} Gray, R.~O., \& Kaye, A.~B.\ 1999, \aj, 118, 2993 

\bibitem[Grigorieva et al.(2007)]{grigorieva07} Grigorieva, A.,
  Th{\'e}bault, P., Artymowicz, P., \& Brandeker, A.\ 2007, \aap, 475,
  755  

\bibitem[Hillenbrand et al.(2008)]{hillenbrand08} Hillenbrand, L.~A., 
et al.\ 2008, \apj, 677, 630 

\bibitem[Holland et al.(1998)]{holland98} Holland, W.~S., et al.\ 1998, \nat, 392, 788 

\bibitem[Houck et al.(2004)]{houck04} Houck, J.~R., et al.\ 2004,
\apjs, 154, 18

\bibitem[Jura et al.(1998)]{jura98} Jura, M., Malkan, M., White, R., Telesco, C., Pina, R., \& Fisher, R.~S.\ 1998, \apj, 505, 897 


\bibitem[Kalas et al.(2008)]{kalas08} Kalas, P., et al.\ 2008,
  Science, 322, 1345  

\bibitem[Kamp \& Paunzen(2002)]{kamp02} Kamp, I., \& Paunzen, E.\ 2002, \mnras, 335, L45 

\bibitem[Kenyon \& Bromley(2004)]{kenyon04} Kenyon, S.~J., \& Bromley,
B.~C.\ 2004, \apjl, 602, L133


\bibitem[Kenyon \& Bromley(2008)]{kenyon08} Kenyon, S.~J., \& Bromley, B.~C.\ 2008, \apjs, 179, 451 

\bibitem[Krist (2006)]{krist06} Krist, J. E. 2006, Spitzer Tiny TIM User’s Guide Version 2.0

\bibitem[Lafreni{\`e}re et al.(2009)]{lafreniere09} Lafreni{\`e}re, 
D., Marois, C., Doyon, R., \& Barman, T.\ 2009, \apjl, 694, L148 

\bibitem[Lagrange et al.(2008)]{lagrange08} Lagrange, A.~-., et al.\
2008, arXiv:0811.3583

\bibitem[Laor \& Draine(1993)]{laor93} Laor, A.~\& Draine, B.~T.\
1993, \apj, 402, 441

\bibitem[Lowrance et al.(2005)]{lowarance05} Lowrance, P.~J., et al.\ 2005, \aj, 130, 1845 

\bibitem[Lu et al.(2008)]{lu08} Lu, N., et al.\ 2008, \pasp, 120, 328


\bibitem[Magnani et al.(2000)]{magnani00} Magnani, L., Hartmann, 
D., Holcomb, S.~L., Smith, L.~E., \& Thaddeus, P.\ 2000, \apj, 535,
167

\bibitem[Malhotra(1995)]{malhotra95} Malhotra, R.\ 1995, \aj,
  110, 420

\bibitem[Marois et al.(2008)]{marois08} Marois, C., Macintosh, B.,
  Barman, T., Zuckerman, B., Song, I., Patience, J., Lafreni{\`e}re,
  D., \& Doyon, R.\ 2008, Science, 322, 1348 

\bibitem[Mart{\'{\i}}nez-Galarza et al.(2009)]{martinez09} 
Mart{\'{\i}}nez-Galarza, J.~R., Kamp, I., Su, K.~Y.~L., G{\'a}sp{\'a}r, A., 
Rieke, G., \& Mamajek, E.~E.\ 2009, \apj, 694, 165 


\bibitem[Meyer et al.(2006)]{meyer06} Meyer, M.~R., Backman, D.~E.,
Weinberger, A.~J., \& Wyatt, M.~C.\ 2007, in Protostars and Planets V,
(Tucson: Univ. Arizona Press), 57


\bibitem[Mo{\'o}r et al.(2006)]{moor06} Mo{\'o}r, A., {\'A}brah{\'a}m, P., Derekas, A., Kiss, C., Kiss, L.~L., Apai, D., Grady, 
C., \& Henning, T.\ 2006, \apj, 644, 525 

\bibitem[Malhotra(1998)]{malhotra98} Malhotra, R.\ 1998, Solar 
System Formation and Evolution, ASP Conference Series, 149, 37 

\bibitem[Moerchen et al.(2007)]{moerchen07} Moerchen, M.~M., Telesco, C.~M., De Buizer, J.~M., Packham, C., \& Radomski, J.~T.\ 2007, \apjl, 666, L109 


\bibitem[Morales et al.(2009)]{morales09}Morales, F.~Y., et al.\ 
2009, \apj, 699, 1067

\bibitem[Moro-Mart{\'{\i}}n \& Malhotra(2005)]{amaya05} Moro-Mart{\'{\i}}n, A., \&
Malhotra, R.\ 2005, \apj, 633, 1150 


\bibitem[Podosek \& Ozima(2000)]{podosek00} Podosek, F.~A., \& Ozima, M.\ 2000, Origin of the earth and moon, edited by R.M.~Canup and K.~Righter and 69 collaborating authors.~Tucson: University of Arizona Press., p.63-72, 63 


\bibitem[Quillen(2006)]{quillen06} Quillen, A.~C.\ 2006, \mnras, 
372, L14 

\bibitem[Reidemeister et al.(2009)]{reidemeister09} Reidemeister, M.,
  et al.\ 2009, \aap, in press. 

\bibitem[Rhee et al.(2008)]{rhee08}Rhee, J. H., Song, I., \& Zuckerman, B. 2008, \apj, 675, 777

\bibitem[Rieke et al.(2004)]{rieke04} Rieke, G.~H., et al.\ 2004,
\apjs, 154, 25

\bibitem[Royer et al.(2007)]{royer07} Royer, F., Zorec, J., \& G{\'o}mez,
A.~E.\ 2007, \aap, 463, 671 

\bibitem[Sadakane, K. (2006)]{sad06}Sadakane, K. 2006, PASJ, 58, 1023

\bibitem[Smith et al.(2009)]{smith09} Smith, R., Churcher, L.~J., Wyatt, M.~C., Moerchen, M.~M., \& Telesco, C.~M.\ 2009, \aap, 493, 299 

\bibitem[Song et al.(2005)]{song05}Song, I., Zuckerman, B., Weinberger, A. J., \& Becklin, E. E. 2005, Nature, 436, 363

\bibitem[Stansberry et al.(2007)]{stansberry07} Stansberry, J.~A., et
al.\ 2007, \pasp, 119, 1038

\bibitem[Stapelfeldt et al.(2004)]{stapelfeldt04} Stapelfeldt, K.~R., et al.\ 2004, \apjs, 154, 458 
\bibitem[Strubbe \& Chiang(2006)]{strubbe06} Strubbe, L.~E., \& Chiang, E.~I.\ 2006, \apj, 648, 652 



\bibitem[Su et al.(2005)]{su05} Su, K.~Y.~L., et al.\ 2005, \apj, 628,
487

\bibitem[Su et al.(2006)]{su06} Su, K.~Y.~L., et al.\ 2006, \apj, 653,
675

\bibitem[Su et al.(2008)]{su08} Su, K.~Y.~L., et al.\ 2008, \apjl,
679, 125

\bibitem[Sylvester et al.(1996)]{sylvester96} Sylvester, R.~J., Skinner, C.~J., Barlow, M.~J., \& Mannings, V.\ 1996, \mnras, 279, 915 

\bibitem[Turcotte \& Charbonneau(1993)]{turcotte93} Turcotte, S., \& Charbonneau, P.\ 1993, \apj, 413, 376 

\bibitem[van Dishoeck(1992)]{vandishoeck92} van Dishoeck, E.~F.\ 
1992, Chemistry and Spectroscopy of Interstellar Molecules, 69

\bibitem[van Leeuwen(2007)]{vanleeuwen07} van Leeuwen, F.\  2007, \aap, 474, 653 

\bibitem[Wahhaj et al.(2005)]{wahhaj05} Wahhaj, Z., Koerner, D. W., Backman, D. E., Werner, M. W.,
Serabyn, E., Ressler, M. E., \& Lis, D. C. 2005, \apj, 618, 385


\bibitem[Werner et al.(2004)]{werner04} Werner, M.~W., et al.\ 2004, \apjs, 154, 1 

\bibitem[Williams \& Andrews(2006)]{williams06} Williams, J.~P., \& Andrews, S.~M.\ 2006, \apj, 653, 1480 

\bibitem[Wetherill(1992)]{wetherill92} Wetherill, G.~W.\ 1992, Icarus, 100, 307 

\bibitem[Wolf \& Hillenbrand(2003)]{wolf03} Wolf, S., \& Hillenbrand, L.~A.\ 2003, \apj, 596, 603 


\bibitem[Wyatt et al.(2007)]{wyatt07} Wyatt, M.~C., Smith, R., 
Su, K.~Y.~L., Rieke, G.~H., Greaves, J.~S., Beichman, C.~A., 
\& Bryden, G.\ 2007, \apj, 663, 365 

\bibitem[Wyatt(2008)]{wyatt08} Wyatt, M.~C.\ 2008, \araa, 46, 339 

\bibitem[Zubko et al.(1996)]{zubko96} Zubko, V.~G., Mennella, 
V., Colangeli, L., \& Bussoletti, E.\ 1996, \mnras, 282, 1321 

\bibitem[Zuckerman \& Song(2004)]{zuckerman04} Zuckerman, B., \& Song, I.\ 2004, \apj, 603, 738 

\end{thebibliography}
\end{document}